# THE BUILDING OF ONLINE COMMUNITIES:
## AN APPROACH FOR LEARNING ORGANIZATIONS, WITH A PARTICULAR FOCUS ON THE MUSEUM SECTOR

Alpay Beler[1], Ann Borda[1], Jonathan P. Bowen[2] and Silvia Filippini-Fantoni[3]

[1] The Science Museum, Exhibition Road, South Kensington, London SW7 2DD, UK
ann.borda@nmsi.ac.uk   http://www.sciencemuseum.org.uk
[2] London South Bank University, Institute for Computing Research, Faculty of BCIM
Borough Road, London SE1 0AA, UK
jonathan.bowen@lsbu.ac.uk   http://www.jpbowen.com
[3] Université Paris I-Sorbonne, Centre de Recherche Images et Analyses, Paris, France
Silvia.Filippini-Fantoni@malix.univ-paris1.fr   http://imagescognitions.univ-paris1.fr

**Abstract**

This paper considers the move toward and potential of building online communities, with a particular focus on the museum sector. For instance, the increase in the use of 'personalized' toolkits that are becoming an integral part of the online presence for learning organizations, like museums, can provide a basis for creating and sustaining communities. A set of case studies further illustrates working examples of the ways in which personalization and specific tools are developing collaborative spaces, community channels and group interactions.

## INTRODUCTION

Virtual communities (Rheingold, 2000) have now been established for decades (e.g., the legendary WELL). However they continue to develop and expand in potentially unexpected ways (Kolko, 2003). Witness, for example, the popularity of 'weblogs' (or 'blogs' for short) as a means to make online diaries available to a wide readership with little technical expertise required. However, technological skills and advancing dynamic web technology can be used to facilitate in the development of community-oriented web resources (Powazek, 2001). Although initially small in size, larger virtual communities based around significant data resources have now been dubbed 'information cities' (Sairmesh *et al.*, 2004).

Increasingly, learning organizations have an online presence but not all are moving beyond the web as a means of publishing organizational information. Museums, for example, are in an ideal position to focus on moving towards using the Internet to develop outreach and a web-based relationship with individuals, communities or sectors. This enables a continuing dialogue with users, a closer-relationship with audiences and user bases, and is importantly a way of bringing together specific groups and knowledge chains across boundaries (e.g., sectoral or geographical).

At its simplest, there are growing examples of organizations from local history societies to national museums that are creating complementary physical and virtual spaces, such as providing online access to object collections and features from exhibitions and special events. This not only raises profile and interest, but also extends the 'mission' of the organization and its existing community base.

At a more involved level, the networking platform that the Internet offers can reach beyond individual organizations to form virtual consortia. For example, the Virtual Museums of Canada (VMC) [www.virtualmuseum.ca] serves as a national aggregator of content and links and acts as a hub for the entire Canadian cultural heritage community. As a platform of selected services, the VMC can also ensure that the smaller, 'link-poor' sites that contain immeasurably valuable local information can be part of this knowledge network and increase 'visitorship' beyond its geographical confines.

At whichever level of virtuality, what is shared is a recognition that communities are the traditional building blocks of culture (OTA, 1990, p. 188). Indeed, the active online participation and engagement of community members can be seen as essential to cultural stewardship, not unlike the way communities are working in academic forums; see, for instance, the H-Net Humanities and Social Sciences Online portal [www.h-net.org]. The creation of a virtual community can also serve the integration of cultural heritage into the popular consciousness, although its ultimate success will derive from the community's ability to meet the real needs of its members. Electronic mailing lists such as Museum-L, H-Museum and other more specialist lists are helping to form online communities of museum professionals, as reported and surveyed in Bowen *et al.* (2003).

The idea of a community-oriented online museum has been accepted at a research level for a while (Friedlander, 1998), and indeed museum organizations recognized the need to address communities too (Williams & Thomas 1997), but we believe that the time has now come to put this into practice more widely with better online support.

There are already various tools that support the aggregation of individuals and the virtual community building process in general. In particular, these tools have the potential to extend the function of the 'online' museum so that it is not just a catalogue of information. Indeed, museums should not just be seen as custodians of knowledge, but part of a wider exchange. As a consequence, community-centric and collaborative activities that might be championed by museums could be undertaken using the following Internet tools:

- Online discussions
- Online seminars
- Chat forums
- Surveys/polls
- Webcasts
- Message boards
- E-newsletters
- Personalization

Significantly, these channels can lend themselves to the development of communities within the museum sector and beyond in which groups of stakeholders can come together due to shared interests in a topic or an activity.

## PERSONALIZATION

Of special note is the increasing use of 'personalized' toolkits that can draw from content/knowledge aggregators like museums and from user-led contributions based around topical themes, debates, forums, etc.

In the past few years, the number of people visiting museum websites has rapidly increased. As a consequence, museums have to face the challenge of creating virtual environments that are increasingly adapted towards the different needs, interests and expectations of their heterogeneous users. One of the solutions available is the introduction of personalization techniques that, by providing differentiated access to information and services according to the user's profile, make facilities, applications and content more relevant and useful for individual users.

By facilitating web navigation and helping people find the right information, at the best level of detail, personalization has enormous potential in the museum world: it can improve the learning process, stimulate visitors' loyalty, attract new audiences and last, but not least, contribute to the creation and development of online communities. This community aspect, which has often been overlooked by professionals, is becoming progressively more important for museums. In fact, thanks to personalized applications such as alerts, thematic newsletters, customizable calendars and recommendation systems[1], that provide tailored content to people with specific interests, museums can easily identify homogeneous communities of users with the same concerns and needs.

Once these different online communities have been identified, it is in the museum's interest to make them operational by developing tools and services that enable their proper functioning, especially by stimulating communication. This is when personalization once again comes into play. Museums can in fact offer a selection of tools on their websites that allow their users to save images, articles, links, search results, as well as other types of information during navigation. By doing so, the user creates a *personal environment* within the museum's website, where they can come back and find information most likely to be of interest, to which new items could be continuously added. The space can be further expanded with other personalized services such as personalized agendas, e-cards and personal galleries/exhibitions.

These types of application are mainly conceived for specialist communities of visitors that use the museum's website as a proper working tool, such as teachers, journalists, experts, students or researchers. In particular, the value of these applications for certain categories of users such as students and teachers can be particularly high. The personal environment can, in fact, offer educators the possibility to make suggestions of exhibits for their students to visit and questions that they would like them to answer during the exploration. In response, the students can save links to the exhibits that most interest them, as well as making short notes both about questions they had at the beginning and about new questions that arise during the exploration (Bowen & Filippini-Fantoni, 2004).

If personal environments offer online communities the possibility to function and grow, *online forums* can further improve their performances by facilitating communication and exchange among members on topics proposed both by users and by curators. In particular, online forums could benefit from the introduction of personalizing features such as notification of debates or issues that might be of interest to the user, information about other users with interests on specified topics (facilitating the networking between community users), personalized news generation based on personal interests, etc. These kinds of personalized services can increase the value of the underlying 'e-community' beyond a social networking environment: 'the community website becomes an attractive permanent home base for the individual rather than a detached place to go online to socialize or network, thus strengthening the relation between the user and the institution' (Case *et al.*, 2003).

---

[1] These applications are currently available on a number of different museums' websites such as the Metropolitan Museum of Art, the Whitney Museum of American art, etc. For a detailed description of these applications, see Bowen & Filippini-Fantoni, 2004.

As underlined above, the role that personalization can play in identifying, stimulating and connecting museums online communities has potential importance. However, so far only a limited numbers of museums have been able to put these principles into practice. A notable example is the *Ingenious* project, being undertaken in the UK National Museum of Science and Industry group (see case studies later), which includes both an online forum facility and the possibility of creating a personal environment.

The *Louvre* museum's proposed new website is also intended to follow a similar approach for the development of online communities through the use of personalized services (Filippini-Fantoni, 2003). The first version of the new site, which is scheduled to be online at the beginning of 2005, will offer a set of personalized services such as alerts, a personalized agenda, personal e-cards, recommendations and thematic newsletters with the aim of creating a group of online communities. Their specific needs and interests will be addressed later in the year when a second version of the new site will be available that is intended to include further personalized services that will help the communities to become operational. A selection of online forums and personalization of the navigation as well as the creation of environments dedicated to specific communities of users such as children, young people, experts, journalists, companies, etc., are planned. Users will be able to find services that have been specifically conceived for them.

Even if such a comprehensive approach has been adopted so far only by the Louvre and the *Ingenious* project in the museum sector so far, contributions by other museums are also worth mentioning. For example the *Cité de Sciences et de l'Industrie* in Paris has developed a system called *Visite Plus* that allows people not only to follow up on their visit using a personal page on the museum's website, in which the path followed as well as a results of their interaction with the museum's devices are displayed, but also to register for a personalized periodical newsletter that focuses on a series of themes selected by the visitors, at the moment of the registration, from a list of available subjects. Since the system was introduced, the response has been quite positive (Bowen & Filippini-Fantoni, 2004).

Roland Topalian, multimedia developer of the Visite Plus system, has noted that these applications, the personalized newsletter in particular, have contributed to the creation of a strong community of users that often come back to the museum to become involved in its activities. A similar approach is also followed by other museums, like the *Metropolitan Museum of Art* ('Met'), which by introducing a series of personalized applications such as 'My Met calendar', 'My Met museum', a 'Remind me' option and a 'Send this to a friend' service, has been among the first to realise the potential that personalization has in both creating and sustaining online communities.

## CASE STUDIES

A selection of examples in which online tools can be seen as assisting in the development of a community-base by museums and related organizations are presented in this section.

## (1) *Ingenious* – a 'knowledge' site of virtual collections, narratives and personal 'create' tools

The Science Museum in London undertakes many significant IT-based projects to enable networked access to science resources (Borda & Bud, 2002; Borda & Beler, 2003). *Ingenious* [www.ingenious.org.uk]*,* funded by the NOF-digitise programme (a UK Lottery-funded initiative), represents the largest scale object-based Internet project to date undertaken by the museum. It will make publicly accessible 30,000 digitized images and accompanying records, 10,000 library records, and 10,000 object records sourced from the Science Museum (including the Science and Society Picture Library) and its sister sites, the National Railway Museum in York and the National Museum of Photography, Film & Television in Bradford.

In addition, this material is contextualized by subject-related debate forums and several hundred pages of about forty topical stories aimed at life-long learners. These topics use the primary material 'to weave connections between the people, innovations and ideas that have changed our lives and the way we see the world, from the industrial revolution to the present day' (from the *Vision statement*). Through these connections, users have the further opportunity to find meaning for themselves by being presented with tools to create and contribute to subject-driven debates.

The creation of a personalized environment is one of the imperative aims of *Ingenious*. By providing the user with customizable tools to create their own stories, for example, the organization can start directing the growth of a community through topics and by engaging the interests of its members. Topics also define what can exist in a community, as well as providing a type of social ontology. Similarly, debate contributions offer a richer dynamic to a thread and can become an important form of group activity that can potentially encompass a greater cross-section of audience than is possible for similar in-house events.

There are five customizable tools available on the site to users who register (see Table 1 below). If the user registers on the site, they can also allow others to view and share their 'personal' space and have the ability to send any of their personal resources via email. Additionally, registered users have access to the debate forums and can contribute and join conversations focused around topics such as the human genome project and other contemporary issues.

| Tool | Description of Features |
|---|---|
| **My 'links'** | Users can add a copy of a link to their personal links area as they are browsing through the site. Within the 'my links' section, links can be modified or appended with a brief description. A text file of the links can be sent as email or downloaded as a text file. |
| **My 'search'** | Users can save search queries seamlessly to the 'my search' area and have the option to delete or modify searches. In addition, the user can search directly from their personalized search area. |
| **My 'saved images'** | Users can view the images that they have searched or saved from the site's virtual collection of 30,000. E-cards can be produced from this feature and personalized text added to the image. |
| **My 'web gallery'** | Users can add text and captions to images of objects, to arrange them like a thematic slide show, and to save other types of resources. |
| **My 'debate'** | Users can access the debate forums and can contribute and join conversations focused around topics, and to save links to their contribution or a specific debate area. |

**Table 1: List of *Ingenious* Tools**

A highlight among the personalized tools is the *My 'web gallery'* function. An illustrated story or topical learning resource can be built using this feature, much like the 'webquest' concept based around enquiry-oriented activities whereby users create web pages directly from Internet resources. Furthermore there is the option to send and share a gallery with colleagues, family and friends. In the future plans of the site, web galleries will be published on the *Ingenious* site to permit more users to visibly contribute to and share the knowledge base and to form communities specific to different strands.

Importantly, the *Ingenious* tools and site resources in general enable the support of three key elements that communities also share, namely:

- **Interests:** What interests us and how we feel about it (e.g., web gallery, debates)
- **Information:** Where to find something and how to do something (e.g., searching functions, my links, save image functions).
- **Knowledge:** Opinions, learning and relationships. (e.g., debates, web gallery, save image, site subjects and topics)

In this way, the tools available to users on the site are not just tailored for the individual. The changing scope of personalization has been taken into consideration whereby information seeking and capture can be disseminated, shared and extended by others' contributions, resources and personal knowledge base. This is in line with the movement of personalization becoming a case of 'my community', rather than 'my information' (Rheingold, 2000). As expounded by Rheingold, more users are asking about what they are missing and whether other users think this information is important.

## (2) *Science, Invention and Nature* – a consortium-led search portal

Expanding the community-focus online, the Science Museum is currently leading a consortium project (funded by the NOF-digitise programme) called *Science, Invention and Nature* (SIN) to launch in late summer of 2004 [www.sinergies.org.uk]. SIN will be an interdisciplinary web portal that combines information from four institutional websites and their respective NOF-funded projects exploring aspects of the natural and manmade worlds. The four partner websites (with NOF projects in italics) linked to the portal thus far are:

- Science Museum [www.sciencemuseum.org.uk]
  – *Ingenious* [www.ingenious.org.uk]
- Natural History Museum (NHM) [www.nhm.ac.uk]
  – *Nature Navigator* [www.nhm.ac.uk/naturenavigator]
- Wildscreen Trust (WS) [www.wildscreen.org.uk]
  – *Arkive* [www.arkive.org]
- Y Touring (YT) [www.ytouring.org.uk]
  – *Genetic Futures* [www.geneticfutures.com]

Internet users entering the SIN website can browse through different themes (also known as 'SINergies') and click on related links which will also take the user to content on the four sites. The SIN web site's main task is to collate select content from each of the sites into these broad editorially driven SINergies. One example of such a theme on the SIN portal is 'Food'. The SIN portal will hold introductory text on this theme, as well as food-relevant links to the other four web sites.

There is a possibility for the user to 'drill' the other way as well. The four web sites will each hold a SINergy 'button' on different pages where theme-relevant content will take the user to the SIN portal. For example, a page on 'Eating disorders' on the Y-Touring web site will hold that SINergy button; when a user clicks it, they will be taken to the SIN portal and the theme on 'Food'. The theme will then lead to further content on the topic of 'Food' through hyperlinks and a relevant keyword search (pre-defined, based on the subject page on which the button resides).

This concept follows the idea of a 'webring' in which a group of websites with a common theme, linked in a loop, allows the user easy access in the ring by clicking on navigational links. It also enables users to interact with other groups or communities within the ring. Thus, a user may be introduced to a community already part of one of the partner sites or cross-site groups. On the main site itself, there is a mechanism to generate polls related to select themes; this will provide a basic means of starting a topical community.

To support the concept of an 'enclosure' of subject-specific resources, the other principle component of the SIN web portal is a sophisticated search engine. One of the main purposes of the portal is to provide users with the possibility to search for content concerning relevant topics and to provide them with a 'complete' result across all underlying partner sites via search algorithms and metadata. The user can also search by resource 'type' (e.g., image, PDF, HTML page, etc.).

The search is a particularly key component because SIN represents a resource almost entirely comprising of 'born digital' resources (i.e., website materials), and the site itself has minimal content; the SINergies act as 'jumping off' points and a means to provide examples of the range of interdisciplinary content that is available.

Essentially, via these features of the SIN site, there is the potential to nurture interest groups and communities linked to the subject hierarchies of the site. As in the *Ingenious* site, the direction of a community is intrinsically based on what topical subjects exist. These engage the interests of the members and support community management by providing a tangible social ontology.

The idea behind the overall concept is to offer users something more targeted than a Google result set; here users are provided with well-known and branded knowledge organizations to search across and access authoritative learning materials. The next step in the community-building process is to provide more tools on the site to create resources from the available content and to allow more 'informal' knowledge to be contributed to the existing themes.

## (3) *SOAS Digital Media Resource Library (DMRL)* – a digital library supporting a virtual learning environment

During the late 1990's, SOAS (School of Oriental and African Studies), part of the University of London, conducted a survey of the multimedia needs of its constituent departments. One notable outcome of this analysis was to demonstrate the need of greater access to large sets of slide and sound archives that were stored in an analogue format. The result was a pilot project to create an online Digital Media Resources Library (DMRL) [mercury.soas.ac.uk/it/projects/project03], whereby the Art & Archaeology department was chosen as a pilot case.

The project's principle objective was to provide digital media resources for teaching and research to staff and students, in the first instance within the Art & Archaeology department itself via the SOAS intranet, but with a view to making this resource available to the wider academic community in due course. The digitized media included slides, photographs, sound files and video clips.

The overall achievement of the project within a very limited budget was not only to create an accessible cross-departmental digital resource, but to lay the ground for a virtual learning environment via which the course tutors could make available the main body of their teaching materials online for their students to study and to use in the preparation of course work. Additionally, students were able to access the key images necessary for their specialist examinations through the DMRL database.

The database metadata, in particular, were carefully developed in consultation with specialists in the various subject areas so that both students and lecturers could benefit from keyword and more targeted taxonomic queries. One query tool enabled teachers to build extensive sets of database queries that were available for their own students to access and which they could run themselves in order to generate the relevant research material. For a number of subject areas, key course materials were supported with lecture notes together with references to books and other useful external links that were tagged with relevant metadata.

The DMRL aim to build communities for research collaboration and sustainability purposefully enabled PhD students to 'plug-in' to the DMRL infrastructure to digitize and document their rich media, as well as to produce multimedia based PhDs. This was seen as essential for generating and preserving new authoritative resources and creating a socio-technical system that underpinned another form of group interaction and key knowledge capture. PhD students could be part of a 'publishing' process and gain a sense that their research was taking on a number of tangible and accessible forms. These extensive PhD resources were, with their creators' permission, then re-incorporated in the DMRL for use by the wider student and lecturer community.

It should be noted that the functional goals for service delivery to the various target groups drove most of the direction of the DMRL, whereby the social goals that can facilitate virtual interactions and communities were readily supported in the lecture room (Schraeffel, 2000). Hence, the DMRL's community building largely supported a blended style of interaction that drew from a database of tools and user-generated resources and on which the student and teacher community could base their in-class discussions.

For instance, classrooms and lecture theatres were equipped with data projectors, enabling teachers to present a more diverse range of materials accessible from the DMRL than if they were only equipped with slide projectors. This also presented the students with a unique opportunity to participate, allowing them to challenge the course materials and discussions with their own 'discovered' resources from the DMRL.

In general, on an overall organizational level, the DMRL project generated a valuable discourse on how to deliver and expand future e-learning materials for a community beyond the departmental boundaries. Firstly, departments were coaxed into taking take stock of their 'hidden' resources and materials, seeking ways of promoting them and making them 'visible' for research across the organization. Secondly there was an increase in the expectations of students and lecturers in both delivery and dissemination of knowledge and consequential benefits. Thirdly, taught courses could be far more responsive and proactive to the research material needs of students and academics. Tangibly, the Art and Archaeology Department was able to attain 5/5 scoring on the HEFCE Quality Teaching Assessment (QTA) with the inclusion of DMRL within their teaching provision. Plans for the future now include opening the DMRL access from the SOAS intranet, making it available via the Internet to emulate and extend its present model.

## (4) *Museophile* museum discussion forums

The museum professional community is still relatively disjoint online. Some efforts have been made to engender a virtual community spirit for visitors to individual museums (Gaia, 2001) but there is not a great deal of cross-museum support. Much of the content for museum discussion forums is distributed around the Internet in a rather inaccessible manner than is difficult to find, in mailing lists on various servers (Bowen *et al.*, 2003). A useful facility could be one that provides a single gateway website where many of these resources are gathered together in one location. As an experiment, such a website has been established as part of a project by London South Bank University and Museophile Limited. Figure 1 shows an example article on this website.

**Figure 1: Screenshot of Museophile discussion forums website
[forums.museophile.net]**

Museophile was developed as a spinout project from London South Bank University that originally started in September 2001 [www.museophile.com]. As part of this project, the Museophile discussion forums website was founded on 4[th] August 2002 [forums.museophile.net]. This facility allows a customized set of mailing lists and news feeds to be included on the main homepage. The inspiration for this facility grew up of the Virtual Library museum pages (VLmp) that gather a great number of links to general museum-related websites (Bowen, 2002b).

This site has two major features, local forums and a selection of external mailing list, newsgroup and news feeds from other sites. Within the website, there are a number of forums split into named sections that visitors can peruse such as announcements, conferences, events, jobs, news, websites, etc. This list can easily be expanded or amended at any time. It is also possible for visitors to post items to these sections and to add comments to existing items. The user can undertake a keyword search across these sections and also within a particular section. Other facilities include polls for voting on questions and diaries (like weblogs) by individual users.

It is worthwhile for anyone making extensive use of this site to register a user account. This is relatively simple and only requires selection of a username and the provision of a working email address. Information is posted to the email address that allows the user to log into the site after which a new password may be selected and further information provided if desired. While it is possible to browse and even post items without a user account, extra facilities are available to those with such an account.

For example, any items posted will be identified with the user and it is also possible to maintain a personal 'diary' of messages if desired.

Perhaps the most useful feature for registered users is the ability to customize the home page website with external newsfeeds. Normally a selection of the most recent messages posted on forums is displayed. As well as these items, registered users can also select from a wide range of feeds from external mailing lists, newsgroups and news sources relevant to museums as discussed earlier in this paper. Around a hundred sources are available and others can relatively easily be added (and suggested by users). When a particular feed is selected by a user, links to the most recent items available from that source are displayed in a list on the right-hand side of the Museophile discussion forums home page. Typically ten to fifteen items are a good number, but this is user-selectable. Any number of feeds may be displayed, although if many are selected the page could become quite large.

The interface to Museophile discussion forums is meant to be relatively simple and usable on a range of browsers. In addition, a text-based interface with user-selectable size, colours and fonts is available, aimed at those with sight difficulties, is also available [access.museophile.net/forums.museophile.net], helping to make the website accessible to a wider set of users. Museophile is especially interested in improving accessibility to museum websites (Bowen, 2003; Bowen, 2004) both for discussion forums and collaborative e-commerce (Bowen, 2002a) to help improve the community spirit in both these areas.

With the above facilities, a registered Museophile discussion forums user has a single point of access for a wide range of museum-related information sources, from local web forums to mailing lists, newsgroups and news sources. The system is not perfect of course, mainly due to limitations in the underlying software. For example, it is not currently possible to search across the external feeds and posting messages is not as simple as it could be. However it is a useful resource and as of May 2004 it has over 720 registered users. Readers of this paper are welcome to join and experiment with the facilities [forums.museophile.net/newuser].

To help assess the needs of the museum professionals community and to improve the Museophile forum facilities, an associated questionnaire was installed and advertised on some of the major museum electronic mailing lists, selected newsgroups and on the Virtual Library museums pages in 2002. The questionnaire was made available in both English and French and the results from around 150 respondents are reported elsewhere (Bernier & Bowen, 2004).

The underlying technology for the Museophile forums website uses the Resource Description Framework (RDF), a standard endorsed by the World Wide Web Consortium [www.w3.org/RDF] and the associated RDF Site Summary (RSS) format, based on the generic Extensible Markup Language (XML). This simple but powerful technology facilitates the discovery of timely information on the web, in contrast to search engines where access to content is always delayed typically by weeks due to the time taken for the web to scan new data. Google, the leading search engine, provides a web interface with around 4,500 news sources [news.google.com] continuously updated using RDF technology.

There are not many museum-related RSS newsfeeds available yet, but a notable high-quality exception is that provided by the UK 24 Hour Museum (Pratty, 2004) [www.24hourmuseum.org.uk/etc/formuseums/TXT18198_gfx_en.html]. Individual museums could provide newsfeeds of the press releases and other timely announcements. For example the museums at the University of Berkeley California provide six separate newsfeeds. All these and others are available on the Museophile museum forums website.

Any reader who has never subscribed to an online forum may be interest to try the Museophile museum discussion website (see Figure 1). The site won recognition with an Honourable Mention at the 2003 *Museums and the Web* conference Best of the Web awards under the Best Museum Professional's Site category. It is hoped that it will help improve the online community facilities available for museum professionals, particularly for keeping them up to date with developments.

## CONCLUSION

Such scenarios show how virtual communities can be formed to allow instant access to expertise. They can allow collaboration and, in particular, the sharing of best practice and know-how in real-time. They can further provide common ground for solving problems and providing access to resources (including multimedia and datasets). Developer communities have been exemplars in this area since the early days of the Internet. Similarly, those in the education sectors have long built communities around resource sharing and dissemination of tools and information for teaching and learning.

Museums can play a significant role in this direction, especially as they already have experience in both serving a physical community and in developing an online resource base for their virtual counterpart.

Virtual communities are of interest to researchers from a sociological and technical point of view (Hummel and Lechner, 2002). However, the ingredients for success are difficult to assess in a museum setting, with only a little experience so far (Carr, 2002). Bowen and Bernier (2004) presents one of the few surveys of the online museum professional community so far. Some have considered the community involved with individual museums (Gaia, 2001; Hagen-Thofson *et al.*, 2000; Srinivasan, 2003). In other cases, small numbers of museums and related organizations have collaborated (Bazley *et al.*, 2002; Dowden, 2003). An important aspect of museums is their collections, and considering how different communities could access these via a multimedia database, for example, is an important and unique facet of a museum (Shabajee, 2002). One advantage of community involvement is that content may be generated for the museum (Durbin, 2004), although care has to be taken over its quality. In any case, for such communities, it is important that there is a shared purpose, the infrastructure provided is easy to use and that there is a reasonable level of trust both of the technology itself and the other members of the community.

For the future, networking (in the business and social sense) is increasingly being supplemented and aided by web-based tools such as those provided by LinkedIn [www.linkedin.com] and Ecademy [www.ecademy.com] for professional contacts, including museum personnel. These resources enable connections to be recorded and fostered between professionals. Although not many museum personnel are yet participating, LinkedIn includes a 'Museums and Institutions' section and around 160 people have classified themselves under this category as of May 2004. However the highest number of connections by these people to others is around 30, with most less than 10, compared to over 2,500 for the most assiduous networker.

Despite the fact that online museum communities for both visitors and professionals are not yet very well developed in general, we believe the future has good prospects for improvement. The case studies in this paper illustrate a number of important first steps in this direction and we believe that personalization will be an important aspect of most such efforts. The necessary technology is increasingly available and the limitations are likely to be as much political as financial. We look forward to exciting developments in the support of online communities by museums and other learning organizations over the next few years.

**Acknowledgement:** Mike Houghton (Visiting Fellow, London South Bank University, UK) aided in the implementation of the Museophile forums website as part of a London South Bank University BASS (Business Award and Support Scheme) project.